\RequirePackage{fix-cm}

\documentclass{svjour3}                     

\smartqed  
\usepackage{graphicx}
\usepackage[misc]{ifsym}
\usepackage{mathptmx}      

\begin{document}

\title{Primary thermometry in the intermediate Coulomb blockade regime}

\author{A. V. Feshchenko \and M. Meschke \and D.~Gunnarsson \and M. Prunnila \and L. Roschier \and J. S. Penttil\"a \and J.~P.~Pekola}

\institute{A. V. Feshchenko (\Letter)
                 \and M. Meschke \and J.~P.~Pekola \at Low Temperature Laboratory, O. V. Lounasmaa Laboratory, Aalto University, P.O.Box 13500, FI-00076 AALTO, Finland\\
                \email{anna.feshchenko@aalto.fi}\\          
                \and
               D. Gunnarsson  \and M. Prunnila \at VTT Technical Research Centre of Finland, P. O. Box 1000, FIN-02044 VTT, Espoo, Finland\\
               \and 
               L. Roschier \and J. S. Penttil\"a \at Aivon Oy, Valimotie 13A, 00380 Helsinki, Finland\\}

\date{Received: date / Accepted: date}

\maketitle

\begin{abstract}
We investigate Coulomb blockade thermometers (CBT) in an intermediate temperature regime, where measurements with enhanced accuracy are possible due to the increased magnitude of the differential conductance dip. Previous theoretical results show that corrections to the half width and to the depth of the measured conductance dip of a sensor are needed, when leaving the regime of weak Coulomb blockade towards lower temperatures. In the present work, we demonstrate experimentally that the temperature range of a CBT sensor can be extended by employing these corrections without compromising the primary nature or the accuracy of the thermometer.
\end{abstract}

\section{Introduction}
\label{intro}

Since the single electron tunneling phenomena were described about 30 years ago \cite{Averin1986_1}, many devices based on small tunnel junctions have been demonstrated. An example of such a device is the Coulomb blockade thermometer (CBT) using one \cite{Pekola1994} or two dimensional arrays of tunnel junctions \cite{Bergsten1999,Devi2004}. CBT based on electron tunneling enables one to relate the measured voltage to temperature with the help of Boltzmann constant $k_{B}$, which could thus provide a means to revise the definition of kelvin \cite{Fellmuth2006}. A related idea forms the basis of shot-noise thermometry, described in Refs. \cite{Spietz2003,Spietz2006}. 

\section{Theoretical background}
\label{theory}
CBT thermometry is based on the change of electrical conductance $G$ of tunnel junction arrays. Typically, CBT works in a weak Coulomb blockade regime $E_{c} \ll k_{B}T$, where the charging energy of the system with $N$ junctions in series is $E_{c}\equiv[(N-1)/N]e^2/C_{\Sigma}$ and $C_{\Sigma}$ is the total capacitance of an island between the junctions that depends on the physical size of the contact and the self capacitance of the island. In this regime, CBT sensor works as a primary thermometer, meaning that calibration is not needed \cite{Pekola1994}. For a uniform array, the differential conductance scaled by its asymptotic value $G_{T}$ at large positive and negative voltages can be written in terms of bias voltage $V$ as 
\begin{equation} G(V)/G_{T}=1-\frac{E_{c}}{k_{B}T}g\Big(\frac{eV}{Nk_{B}T}\Big),
\label{Eq.1} 
\end{equation}
where the function $g(x)=[x \sinh(x)-4 \sinh^2(x/2)]/[8 \sinh^4(x/2)]$ determines the bias dependence. 

Primary measurements of temperature can be achieved from the measurement of full width of the conductance dip at half minimum 
\begin{equation} V_{1/2} \cong 5.439Nk_{B}T/e. 
\label{Eq.2} 
\end{equation}
Thus, obtained temperature does not depend on the geometry or material of a sensor. A secondary temperature measurement can be obtained by recording the depth of the zero bias conductance dip 
\begin{equation} 
\frac{\Delta G}{G_{T}}=\frac{1}{6}u_{N}, 
\label{Eq.3} 
\end{equation} 
with parameter $u_{N} \equiv E_{c}/k_{B}T$. As was shown in Ref. \cite{Farhangfar1997}, at high temperature (in that case 4.2 K) the measured conductances follow quite well the expression in Eq.~(\ref{Eq.3}).

In this letter, we demonstrate experimentally the extension of Coulomb blockade thermometry into an intermediate temperature regime described theoretically in Refs. \cite{Farhangfar1997,Meschke2004}. Experiments employing quantum dot thermometry \cite{Hoffmann2007,Gasparinetti2011,Gasparinetti2012,Prance2009}, investigating electrical conduction in metallic nanodots \cite{Bitton2011} or granular films \cite{Yajadda2011} work in this intermediate Coulomb blockade regime as well. 

\begin {figure}[h!t]
\centering 
\includegraphics[width=0.6\textwidth]{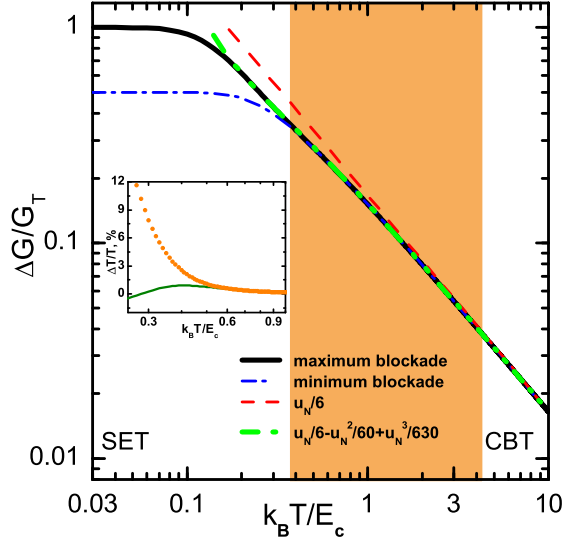} 
\caption{(Color on-line) Theoretical temperature dependence of the depth of the zero bias conductance dip of a tunnel junction array influenced by Coulomb interaction. The left side of the plot represents the working regime of single electron devices (SET), where $E_{c} \gg k_{B}T$. Colored area represents an intermediate regime, where $k_{B}T \sim E_{c}$, which is the focus of this work. On the right side of the plot the normal working regime of CBT sensors is represented, where $E_{c} \ll k_{B}T$. Two extremes of the possible experimental results due to unknown background charge configuration are represented by maximum (solid black line) and minimum (dash - dotted blue line) blockade curves. The third order result, Eq.~(\ref{Eq.5}), is presented by the dash double - dotted green line. The inset represents the error in temperature reading which one makes using the third order approximation of Eq.~(\ref{Eq.5}). The orange dotted curve represents the relative difference between the numerically calculated minimum blockade (dash - dotted blue line, Fig.~\ref{fig:1}) curve and the third order curve (dash double - dotted green line, Fig.~\ref{fig:1}). The solid green line on the inset plot represents the relative difference between the maximum blockade (solid black line) and the third order approximation curve.}
\label{fig:1}
\end{figure}

The intermediate regime is defined by $E_{c}\sim k_{B}T$, between the traditional CBT operation range $E_{c} \ll k_{B}T$, and the full Coulomb blockade $E_{c} \gg k_{B}T$, see Fig.~\ref{fig:1}. Solid black and dash - dotted blue lines correspond to the numerically calculated temperature dependent conductance of CBT at zero bias using theory of single electron tunneling \cite{Pekola1994} (refered to "full theory" in the following text). The two limiting cases, "minimum blockade"~(dash - dotted blue line)~and "maximum blockade" (solid black line), represent the two opposite extremes of the possible background charge configurations.

Thus we compare the two extreme cases to indicate the uncertainty due to the unknown background charge values. As long as the two curves follow each other, the uncertainty due to background charges can be neglected. In the regime $E_{c} \gg k_{B}T$ the charge sensitivity plays a dominant role \cite{FultonDolan1987}, as the measured conductance depends on the randomly fluctuating background charge configuration, see Fig.~\ref{fig:1}. To fully tune the background charges in a large array one should have an equally large number of gate controls. Instead of using elaborate gate control, we therefore limit the use of the thermometer to a regime, where the influence of background charges on temperature reading stays sufficiently small.

The intermediate temperature regime is beneficial for thermometry due to enhanced signal magnitude $\propto~E_{c}/k_{B}T$. In addition, the harmful contribution of nonlinear I-V curves (predominantly due to finite tunnel barrier height) as a function of applied bias voltages is less abundant \cite{Gloos2003}. Another motivation of the present work is to expand the temperature range where a single CBT sensor can be used \cite{Behnia2004}.

To obtain the temperature with high accuracy, one has to make a series expansion up to higher orders in $u_{N}$ towards low temperatures \cite{Farhangfar1997}. In this work, we show experimentally that the temperature of  a CBT sensor can be obtained by employing corrections to Eqs.~(\ref{Eq.2}) and (\ref{Eq.3}) as
\begin{equation} 
V_{1/2}\cong5.439Nk_{B}T(1+0.3921\frac{\Delta G}{G_{T}})/e, 
\label{Eq.4} 
\end{equation}
\begin{equation} 
\frac{\Delta G}{G_{T}}\cong\frac{1}{6}u_{N}-\frac{1}{60}u_{N}^2+\frac{1}{630}u_{N}^3. 
\label{Eq.5} 
\end{equation}
Note that as $V_{1/2}$ and $\Delta G/G_{T}$ are measured quantities, Eq.~(\ref{Eq.4}) maintains the primary nature~\footnote{The prefactor (0.3921) in front of the depth of zero bias conductance dip, ${\Delta G}/{G_{T}}$ in Eq. (4), is the original value and the same as in Ref. \cite{Farhangfar1997}. In Ref. \cite{Meschke2004} there is a typo (0.3992).}. We examine here the series expansion up to the third order, which gives an accuracy of 1.1 \% (see in the following text). Since we assume that the scatter in the resistance of the tunnel junctions can bring a comparable error, we do not consider higher order contributions. 

The solution of Eq.~(\ref{Eq.5}) in terms of conductance is given by
\begin{equation} 
T\cong\frac{E_{c}}{k_{B}}\left[\frac{1}{6}\Big(\frac{\Delta G}{G_{T}}\Big)^{-1}-\frac{1}{10}-\frac{1}{350}\Big(\frac{\Delta G}{G_{T}}\Big)+\frac{27}{875}\Big(\frac{\Delta G}{G_{T}}\Big)^2\right].
\label{Eq.6} 
\end{equation} 

In Fig.~\ref{fig:1}, one can see that Eq.~(\ref{Eq.3}) (dashed red line) at low $k_{B}T/E_{c}$  is no longer valid. The exact dependence of $\Delta G/ G_{T}$ is shown by solid black and dash - dotted blue lines. On the other hand, the higher order series expansion presented by Eq.~(\ref{Eq.5}) (dash double - dotted green line) follows the exact dependence with good accuracy down to lower temperatures. In the inset of Fig.~\ref{fig:1}, the orange dotted curve represents the relative difference between minimum blockade and the third order approximation expressed in Eq.~(\ref{Eq.5}). The solid green curve represents the relative difference between maximum blockade and the third order approximation. To keep the uncertainty in the measurement small, we limit the range of CBT operation to above $k_{B}T/E_{c}\sim 0.4$. This means that the third order result deviates less than 2.5\% from either of the opposite extremes (maximum and minimum blockade) of possible experimental results. Note that although the expansion parameter $u_{N}$ assumes values exceeding unity, the series of Eq.~(\ref{Eq.5}) does not diverge even in this case because the multipliers diminish quickly for higher orders~\footnote{A more natural expansion parameter would be the depth in the lowest order $w_{N} \equiv u_{N}/6$. In this case Eq.~(\ref{Eq.5}) would read $\Delta G/G_{T}=w_{N}- (3/5)w_{N}^{2}+(12/35)w_{N}^{3}$}. 

\section{Experimental Details}
\label{experiment}

\begin{figure}[h!t]
\centering
\includegraphics[width=0.65\textwidth]{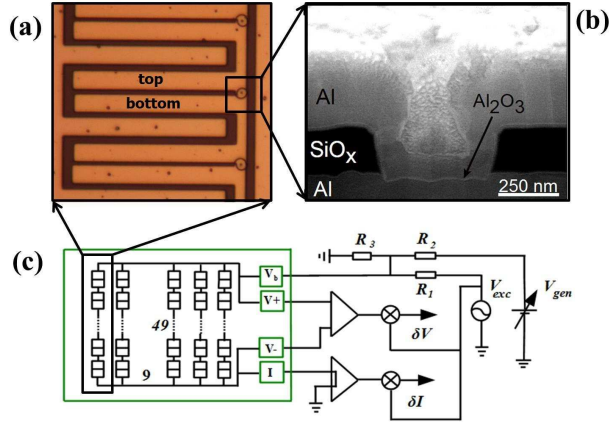}
\caption{(Color on-line) The tunnel junctions and the measurement setup. \textbf{(a)} In an optical microscope image of one of the sensors, three tunnel junctions (circles) formed by overlapping area of two, top and bottom, $Al$  fingers (wide orange lines) are shown. \textbf{(b)} Close-up SEM image of a cross-section of a tunnel junction. The thin $Al_{2}O_{3}$ layer between the two 250 nm-thick $Al$ electrodes is the tunnel barrier. The tunnel junction has round shape with a diameter of 500 nm. \textbf{(c)} Schematic diagram of the measurement setup. A CBT sensor consists of 9 parallel arrays of 49 tunnel junctions in series, which connect to the bonding pads. On the left side of a schematic diagram is the sensor, which is placed in the refrigerator. The right side of a schematic diagram represents the measurement setup at room temperature.}
\label{fig:2}    
\end{figure}

The measured sensors consist of parallel arrays of~tunnel junctions in series. Figure~\ref{fig:2}~(a) shows an image of three junctions.

These sensors were made by ex-situ tunnel junction fabrication method, which utilizes plasma etched via in a dielectric ($SiO_{x}$) for defining the tunnel junctions. Fabrication process is described in detail in Ref. \cite{Prunnila2010}. Tunnel junctions were configured in a round shape of 500 nm in diameter (see Fig.~\ref{fig:2}~(b)). In order to keep the size of the tunnel junctions identical for sensors with different $E_{c}$, the charging energy of a sensor was tuned by varying the overlap area of the cooling fingers. In Fig.~\ref{fig:2}~(a), for the sensor with the highest charging energy the overlap area is very small and noticeable only around the junction area. For the sensor with the lowest charging energy, the top and the bottom cooling fingers are totally overlapping (not shown).

The sensors were measured in a $^3He-^4He$  dilution refrigerator. Detailed schematic diagram of the measurement setup is presented in Fig.~\ref{fig:2}~(c). Differential conductance was measured, as shown in Fig.~\ref{fig:2}~(c), using SR 830 Lock-In amplifiers and DL instruments  low noise voltage (1201) and current (1211) preamplifiers. In addition to voltage biasing $V=V_{gen}R_{3}/(R_{2}+R_{3})$, a small ($2 \%$ of $V_{1/2}$) ac voltage was applied. To obtain the differential conductance $G=\delta I/ \delta V$, both the ac voltage $\delta V$ and current $\delta I$ were measured. 

\section{Results and Discussion}
\label{results}

In the measurement we focus on the intermediate regime, $k_{B}T \sim E_{c}$. We performed the measurements for three CBT sensors with different values of $E_{c}=45$~mK (sensor A), $E_{c}=~94$~mK (sensor B) and $E_{c}=129$~mK (sensor C) in a temperature range between $55$~mK and $470$~mK . In case of sensor C we used a low-charging energy CBT sensor as a reference temperature thermometer (sensor REF, $E_{c}=22$~mK). It has a tunnel junction size of 800 nm in diameter and was operating in the linear CBT working regime, where the second and third order effects are small. 

\begin{figure}[h!t]
\centering
\includegraphics[width=0.6\textwidth]{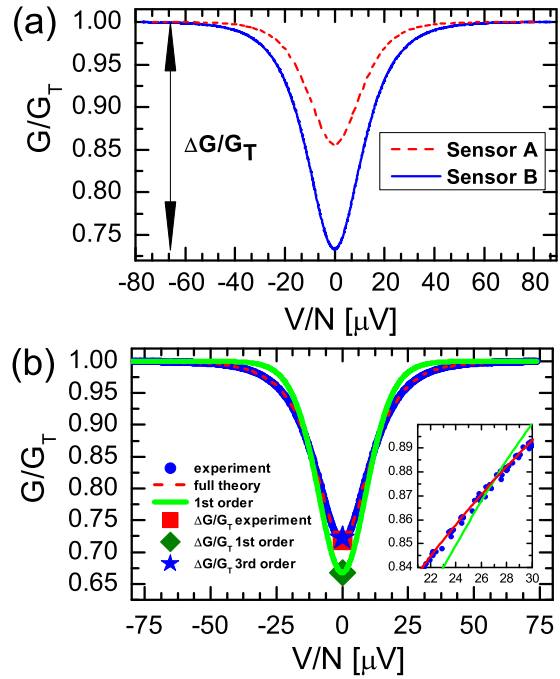}
\caption{(Color on-line) Measured conductance of the CBT sensors at $T=55$~mK determined by $RuO_{2}$ and the theoretical predictions. \textbf{(a)} Normalized conductances $G/G_{T}$ vs applied voltage $V$ measured for sensors A and B. \textbf{(b)} The experimental data for sensor B (blue dots) together with the numerically calculated full theory fit on top (dashed red line). Minimum of the measured conductance at zero bias, which also corresponds to the minimum of the numerically calculated full theory curve, marked as red square, and the minimum of the conductance dip calculated using the third order approximation marked as blue star. The solid green line of Eq.~(\ref{Eq.1}) corresponds to the first order curve and the rhombus to its minimum. The inset shows the full theory curve (red) fitted to measured conductance curve (blue dots) and the first order curve (solid green line) in the bias range near $V_{1/2}/2$.}
\label{fig:3}       
\end{figure}

In Fig.~\ref{fig:3}~(a), we present the measured normalized conductance of sensors A and B vs applied voltage at a refrigerator temperature of 55~mK. The experiment shows the expected difference of the conductance peak depth corresponding to the two charging energies.

In Fig.~\ref{fig:3}~(b), we present the experimental conductance curve (blue dots) for sensor B and the full theory fit to the data (dashed red line). The theoretical model was calculated as follows: first, $I-V$ curves were obtained by employing the master equation, as shown in Ref. \cite{Pekola1994}. In addition, we take overheating effects into account \cite{Meschke2004}. The electron temperature is higher than the lattice temperature, due to the weakness of the electron-phonon interaction. These effects are relatively small due to the cooling fingers (see Fig.~\ref{fig:2}~(a)), which provide a good thermalization between the arrays and the substrate \footnote{At $\pm V_{1/2}/2$, the bias points used in the determination of the temperature, the islands are overheated by about 8~mK at maximum.}. Finally, we differentiate the numerical $I-V$ curves to fit them to the measured conductance curves.

The charging energies and the temperatures were obtained using a standard technique: the sensors were measured at higher refrigerator temperatures (between 420 - 470 mK) and fitted to numerically calculated full conductance curves, as described above. At low temperatures, assuming constant charging energies, we extract the temperatures of the sensors using higher order series expansion expressed in Eq.~(\ref{Eq.4}). During our measurements, we record the temperatures of the fridge via a calibrated $RuO_{2}$ thermometer. At any temperature we have the possibility to compare the $RuO_{2}$ value and the temperature given by the fitting procedure for our sensor. 

The values of $E_{c}=94$~mK and $T=47.1$~mK obtained from the full theory fit to sensor B data in Fig.~\ref{fig:3}~(b) were used to calculate the depth of the conductance dip for the first (Eq.~(\ref{Eq.3})) and the third (Eq.~(\ref{Eq.5})) order approximations. The experimental data yield $\Delta G/G_{T}= 0.2815$, whereas the third order approximation utilizing Eq.~(\ref{Eq.5}) gives $\Delta G/G_{T} = 0.2789$. These values correspond to the red square and the blue star in Fig.~\ref{fig:3}~(b), respectively. As one can see in Fig.~\ref{fig:3}~(b), the numerically calculated full theory curve follows the experimental conductance curve very well, and thus our experiment is in agreement with the theoretical model of single electron tunneling. The value of the depth of the experimental conductance curve (red square) also corresponds to the depth calculated using the third order approximation (blue star), Eq.~(\ref{Eq.5}). The series approximation differs from the result of the full theory by 1.1\% only. The value for the depth of the conductance dip, using only the first order series expansion expressed in Eq.~(\ref{Eq.3}), is $\Delta G/G_{T}= 0.3326$. To show the difference between the first and the third order series expansions, we plot the first order curve (Eq.~(\ref{Eq.1})) on the same graph. The shape of the first order conductance curve is significantly different from the experimental (blue dots) and theoretical (dashed red line) curves. This difference would lead to an error of about 30\% in temperature reading, if the first order result was employed.

\begin{figure}[h!t]
\centering
\includegraphics[width=0.6\textwidth]{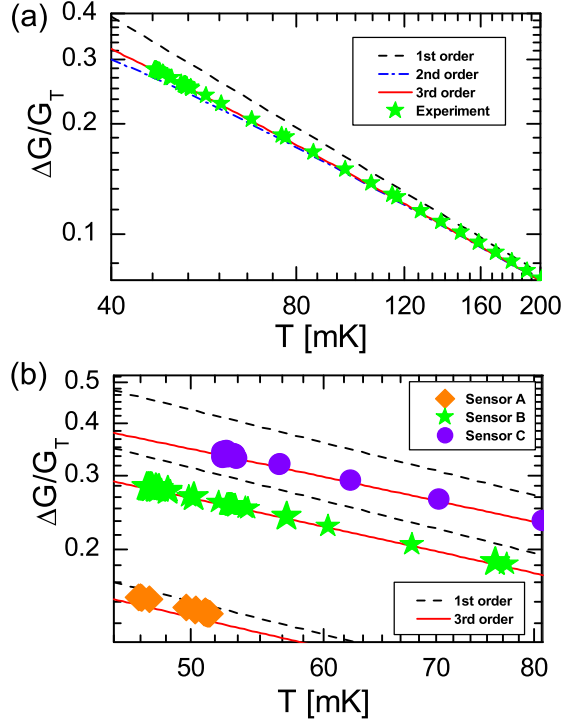}
\caption{\textbf{(a)} The depth of the conductance dip vs temperature for sample B. The three lines correspond to the first (dashed line), the second (dash - doted line) and the third (solid line) order series expansions. \textbf{(b)} Close-up of (a), zoomed to the low temperature regime, for three CBT sensors.  From top to bottom the sets of curves correspond to sensors C, B and A, respectively. The dashed black line corresponds to the first order and solid red line to the third order series expansion. Experimental data are shown as circles, stars and rhombi.}
\label{fig:4}       
\end{figure}

In Fig.~\ref{fig:4}~(a), we present the first, second and third order series expansions (dashed black, dash dotted blue and solid red curves, respectively), calculated using Eq.~(\ref{Eq.5}). On the same plot we show our experimental data points for the depth of the conductance dip for sensor B. The temperature axis is given by the full theory fit to the measured conductance curve. As one can see, the experimental data fit to the third order curve very well at low temperatures, when $E_{c}\cong k_{B}T$. In Fig.~\ref{fig:4}~(b), we present the depth of the conductance curve vs. temperature for sensors A, B and C. We show the experimental data, marked as circles, stars and rhombi, together with the first (dashed black line) and the third (solid red line) order series expansions in 40 - 80 mK temperature range. The data follow the third order curve, calculated using Eq.~(\ref{Eq.5}). 

In Fig.~\ref{fig:4}~(b), the top two curves together with experimental data correspond to sensor C.  We provide an additional proof realizing the temperature scale independently using low-charging energy CBT sensor as reference. Sensor REF, operating in the traditional CBT range, $E_{c} \ll k_{B}T$, was measured at the same time with sensor C. For the latter sensor data points also follow the third order curve. The temperatures of the two sensors agree within 0.5\%. Sensor REF had been earlier compared favorably with the PLTS 2000 temperature scale \cite{Meschke2011} with an accuracy of 1\%. The difference in depths of the conductance dips between the first and the third order series expansions for sensor REF is less than 4\%.

The middle and the bottom sets of curves in Fig.~\ref{fig:4}~(b) correspond to the sensors B and A, respectively. The temperature axis for the two bottom sets of experimental data of these sensors (stars and rhombi) was determined by the full theory fit to our data (as before in Fig.~\ref{fig:4}~(a)). In this case, the fitting forces the depth of the conductance dip to be in accordance to the temperature. We do not observe any deviations of the fitted curve within the experimental accuracy (see inset of Fig.~\ref{fig:3}~(b)). This nearly perfect agreement leaves only a small deviation between the measured depth of the conductance dip and Eq.~(\ref{Eq.5}). The deviation lies within the uncertainty of the value of $E_{c}$, especially when its absolute value becomes smaller, as with sensor A.

\section{Summary}
\label{summary}

To summarize, we demonstrate primary Coulomb blockade thermometry operated in a wide range extending to the $E_{c} \sim k_{B}T$ regime, where the observable signal magnitude and consequently the accuracy are enhanced. We conclude that it is possible to operate the CBT thermometer reliably down to $k_{B}T/E_{c} \cong 0.4$, by employing the third order series expansion. Finally, the accuracy of this method is limited by the influence of background charge distribution, but stays below 2.5\% down to $k_{B}T/E_{c} \cong 0.4$.

\begin{acknowledgements}
We acknowledge the availability of the facilities and technical support by Micronova Nanofabrication Center. We would like to thank A. Peltonen and N. Chekurov for technical assistance. We acknowledge financial support from the European Community FP7 Marie Curie Initial Training Networks Action (ITN) Q-NET 264034 and Tekes through project FinCryo (Grant No. 220/31/2010). This work has been supported in part by the EU 7th Framework Programme (FP7/2007-2013, Grant No. 228464 Microkelvin) and by the Academy of Finland though its LTQ CoE grant (project no. 250280).
\end{acknowledgements}

\end{document}